\PassOptionsToPackage{unicode}{hyperref}
\PassOptionsToPackage{hyphens}{url}
\PassOptionsToPackage{dvipsnames,svgnames,x11names}{xcolor}
\documentclass[
]{article}
\usepackage{amsmath,amssymb}
\usepackage{lmodern}
\usepackage{iftex}
\ifPDFTeX
  \usepackage[T1]{fontenc}
  \usepackage[utf8]{inputenc}
  \usepackage{textcomp} 
\else 
  \usepackage{unicode-math}
  \defaultfontfeatures{Scale=MatchLowercase}
  \defaultfontfeatures[\rmfamily]{Ligatures=TeX,Scale=1}
\fi
\IfFileExists{upquote.sty}{\usepackage{upquote}}{}
\IfFileExists{microtype.sty}{
  \usepackage[]{microtype}
  \UseMicrotypeSet[protrusion]{basicmath} 
}{}
\makeatletter
\@ifundefined{KOMAClassName}{
  \IfFileExists{parskip.sty}{%
    \usepackage{parskip}
  }{
    \setlength{\parindent}{0pt}
    \setlength{\parskip}{6pt plus 2pt minus 1pt}}
}{
  \KOMAoptions{parskip=half}}
\makeatother
\usepackage{xcolor}
\usepackage{longtable,booktabs,array}
\usepackage{calc} 
\usepackage{etoolbox}
\makeatletter
\patchcmd\longtable{\par}{\if@noskipsec\mbox{}\fi\par}{}{}
\makeatother
\IfFileExists{footnotehyper.sty}{\usepackage{footnotehyper}}{\usepackage{footnote}}
\makesavenoteenv{longtable}
\usepackage{graphicx}
\makeatletter
\def\maxwidth{\ifdim\Gin@nat@width>\linewidth\linewidth\else\Gin@nat@width\fi}
\def\maxheight{\ifdim\Gin@nat@height>\textheight\textheight\else\Gin@nat@height\fi}
\makeatother
\setkeys{Gin}{width=\maxwidth,height=\maxheight,keepaspectratio}
\makeatletter
\def\fps@figure{htbp}
\makeatother
\newlength{\cslhangindent}
\newlength{\csllabelwidth}
\newlength{\cslentryspacingunit} 
\newenvironment{CSLReferences}[2] 
 {
  \setlength{\parindent}{0pt}
  \ifodd #1
  \let\oldpar\par
  \def\par{\hangindent=\cslhangindent\oldpar}
  \fi
  \setlength{\parskip}{#2\cslentryspacingunit}
 }%
 {}
\usepackage{calc}

\ifLuaTeX
\usepackage[bidi=basic]{babel}
\else
\usepackage[bidi=default]{babel}
\fi
\babelprovide[main,import]{american}

\def\languageshorthands#1{}
\ifLuaTeX
  \usepackage{selnolig}  
\fi
\IfFileExists{bookmark.sty}{\usepackage{bookmark}}{\usepackage{hyperref}}
\IfFileExists{xurl.sty}{\usepackage{xurl}}{} 
\hypersetup{
  pdftitle={nimCSO: A Nim package for Compositional Space Optimization},
  pdfauthor={Adam M. Krajewski, Arindam Debnath, Wesley F. Reinhart,
Allison M. Beese, Zi-Kui Liu},
  pdflang={en-US},
  colorlinks=true,
  linkcolor={Maroon},
  filecolor={Maroon},
  citecolor={Blue},
  urlcolor={Blue},
  pdfcreator={LaTeX via pandoc}}

\title{nimCSO: A Nim package for Compositional Space Optimization}

\usepackage[affil-it]{authblk}
\usepackage{orcidlink}
\author[1%
  \ensuremath\mathparagraph]{Adam M. Krajewski%
    \,\orcidlink{0000-0002-2266-0099}\,%
    }
\author[1%
  ]{Arindam Debnath%
    \,\orcidlink{0000-0001-9427-4499}\,%
    }
\author[1,2%
  ]{Wesley F. Reinhart%
    \,\orcidlink{0000-0001-7256-2123}\,%
    }
\author[1%
  ]{Allison M. Beese%
    \,\orcidlink{0000-0002-7022-3387}\,%
    }
\author[1%
  ]{Zi-Kui Liu%
    \,\orcidlink{0000-0003-3346-3696}\,%
    }

\affil[1]{Department of Materials Science and Engineering, The
Pennsylvania State University, USA}
\affil[2]{Institute for Computational and Data Sciences, The
Pennsylvania State University, USA}
\affil[$\mathparagraph$]{Corresponding author}
\date{March 4th 2024}

\usepackage[margin=1.05in]{geometry}
\usepackage[scaled]{beramono}
 
\definecolor{darkgreen}{rgb}{0.05, 0.3, 0.1}
\let\oldtexttt\texttt
\renewcommand{\texttt}[1]{\oldtexttt{\textcolor{darkgreen}{#1}}}
\usepackage{caption}

\begin{document}
\maketitle

\hypertarget{summary}{%
\section{Summary}\label{summary}}

\texttt{nimCSO} is a high-performance tool implementing several methods
for selecting components (data dimensions) in compositional datasets,
which optimize the data availability and density for applications such
as machine learning. Making said choice is a combinatorically hard
problem for complex compositions existing in highly dimensional spaces
due to the interdependency of components being present. Such spaces are
encountered, for instance, in materials science, where datasets on
Compositionally Complex Materials (CCMs) often span 20-45 chemical
elements, 5-10 processing types, and several temperature regimes, for up
to 60 total data dimensions.

At its core, \texttt{nimCSO} leverages the metaprogramming ability of
the Nim language (\protect\hyperlink{ref-Rumpf2023}{Rumpf, 2023}) to
optimize itself at the compile time, both in terms of speed and memory
handling, to the specific problem statement and dataset at hand based on
a human-readable configuration file. As demonstrated in the
\protect\hyperlink{methods-and-performance}{Methods and Performance}
section, \texttt{nimCSO} reaches the physical limits of the hardware (L1
cache latency) and can outperform an efficient native Python
implementation over 400 times in terms of speed and 50 times in terms of
memory usage (\emph{not} counting interpreter), while also outperforming
NumPy implementation 35 and 17 times, respectively, when checking a
candidate solution.

\texttt{nimCSO} is designed to be both (1) a user-ready tool,
implementing two efficient brute-force approaches (for handling up to 25
dimensions), a custom search algorithm (for up to 40 dimensions), and a
genetic algorithm (for any dimensionality), and (2) a scaffold for
building even more elaborate methods in the future, including heuristics
going beyond data availability. All configuration is done with a simple
human-readable \texttt{YAML} config file and plain text data files,
making it easy to modify the search method and its parameters with no
knowledge of programming and only basic command line skills.

\hypertarget{statement-of-need}{%
\section{Statement of Need}\label{statement-of-need}}

\texttt{nimCSO} is an interdisciplinary tool applicable to any field
where data is composed of a large number of independent components and
their interaction is of interest in a modeling effort, ranging from
market economics, through medicine where drug interactions can have a
significant impact on the treatment, to materials science, where the
composition and processing history are critical to resulting properties.
The latter has been the root motivation for the development of
\texttt{nimCSO} within the \href{https://ultera.org}{ULTERA Project
(ultera.org)} carried under the
\href{https://arpa-e.energy.gov/?q=arpa-e-programs/ultimate}{US DOE
ARPA-E ULTIMATE} program, which aims to develop a new generation of
ultra-high temperature materials for aerospace applications, through
generative machine learning models
(\protect\hyperlink{ref-Debnath2021}{Debnath et al., 2021}) driving
thermodynamic modeling, alloy design, and manufacturing
(\protect\hyperlink{ref-Li2024}{Li et al., 2024}).

One of the most promising materials for such applications are the
aforementioned CCMs and their metal-focused subset of Refractory High
Entropy Alloys (RHEAs) (\protect\hyperlink{ref-Senkov2018}{Senkov et
al., 2018}), which have rapidly grown since first proposed by
(\protect\hyperlink{ref-Cantor2004}{Cantor et al., 2004}) and
(\protect\hyperlink{ref-Yeh2004}{Yeh et al., 2004}). Contrary to most of
the traditional alloys, they contain many chemical elements (typically
4-9) in similar proportions in the hope of thermodynamically stabilizing
the material by increasing its configurational entropy
(\(\Delta S_{conf} = \Sigma_i^N x_i \ln{x_i}\) for ideal mixing of \(N\)
elements with fractions \(x_i\)), which encourages sampling from a large
palette of chemical elements. At the time of writing, the ULTERA
Database is the largest collection of HEA data, containing over 6,300
points manually extracted from almost 550 publications. It covers 37
chemical elements resulting in extremely large compositional spaces
(\protect\hyperlink{ref-Krajewski2024Nimplex}{Krajewski et al., 2024});
thus, it becomes critical to answer questions like \emph{``Which
combination of how many elements will unlock the most expansive and
simultaneously dense dataset?''} which has \(2^{37}-1\) or 137 billion
possible solutions.

Another significant example of intended use is to perform similar
optimizations over large (many millions) datasets of quantum mechanics
calculations spanning 93 chemical elements and accessible through
OPTIMADE API (\protect\hyperlink{ref-Evans2024}{Evans et al., 2024}).

\hypertarget{methods-and-performance}{%
\section{Methods and Performance}\label{methods-and-performance}}

\hypertarget{overview}{%
\subsection{Overview}\label{overview}}

\begin{figure}[h]
\centering
\includegraphics[width=4.75in,height=\textheight]{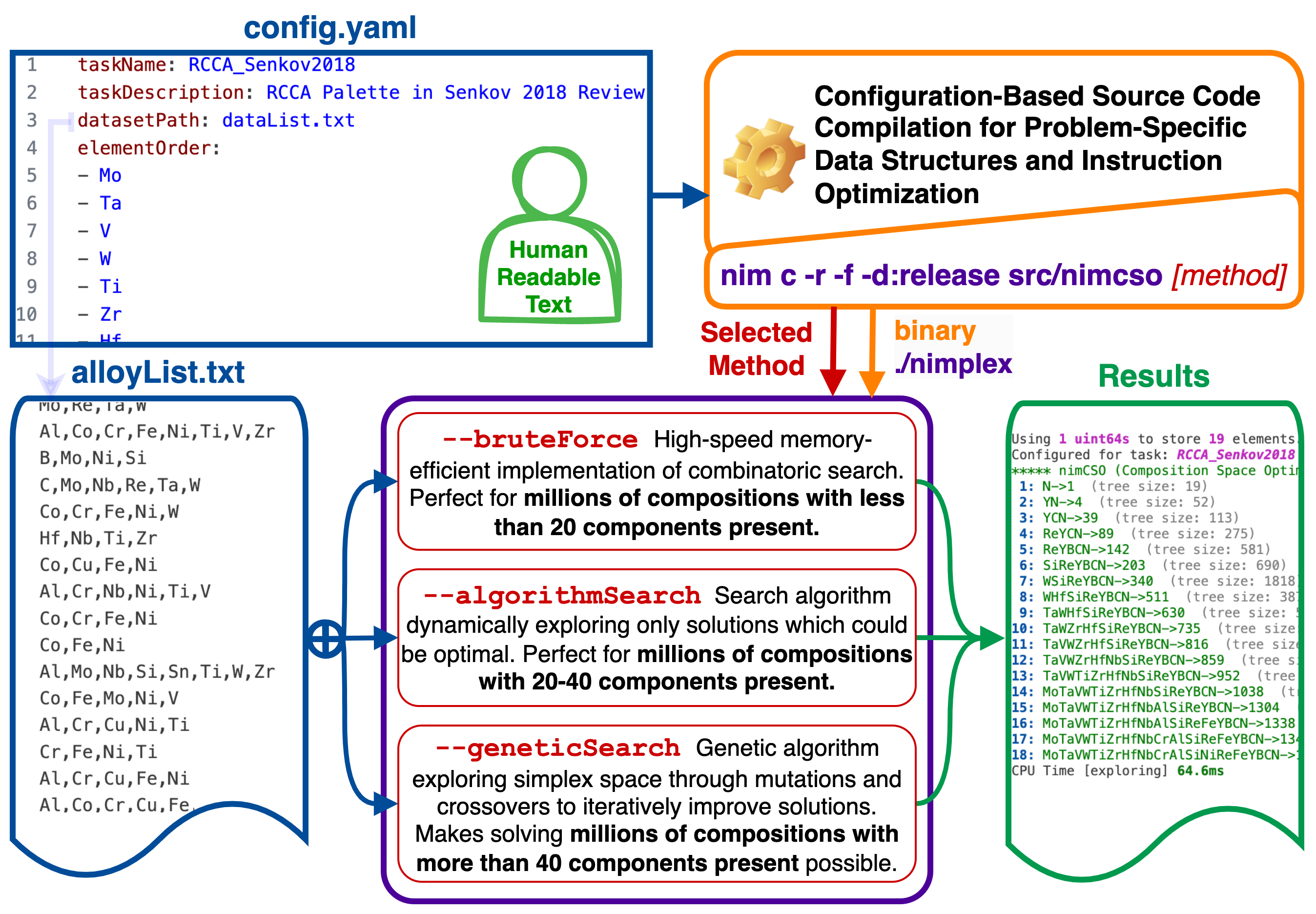}
\caption{Schematic of core nimCSO data flow with a description of key
methods. Metaprogramming is used to compile the software optimized to
the human-readable data and configuration files at
hand.}
\label{fig:main}
\end{figure}

As shown in Figure \ref{fig:main}, \texttt{nimCSO} can be used as a
user-tool based on human-readable configuration and a data file
containing data ``elements'' which can be any strings representing
problem-specific names of, e.g., market stocks, drug names, or chemical
formulas. A single command is then used to recompile
(\texttt{nim\ c\ -f}) and run (\texttt{-r}) problem
(\texttt{-d:configPath=config.yaml}) with \texttt{nimCSO}
(\texttt{src/nimcso}) using one of several methods. Advanced users can
also quickly customize the provided methods with brief scripts using the
\texttt{nimCSO} as a data-centric library.

Internally, \texttt{nimCSO} is built around storing the data and
solutions in one of two ways. The first is as bits inside an integer
(\texttt{uint64}), which allows for the highest speed and lowest memory
consumption possible but is limited to 64 dimensions and does not allow
for easy extension to other use cases; thus, as of publication, it is
used only in a particular \texttt{bruteForceInt} routine. The second
one, used in \texttt{bruteForce}, \texttt{algorithmSearch}, and
\texttt{geneticSearch}, implements a custom easily extensible
\texttt{ElSolution} type containing heuristic value and
\texttt{BitArray} payload, which is defined at compile time based on the
configuration file to minimize necessary overheads. Both encodings
outperform typical native Python and NumPy implementations, as shown in
Table 1.

\captionof{table}{Benchmarks of average time to evaluate how many datapoints
would be lost if 5 selected components were removed from a dataset with
2,150 data points spanning 37 components (10,000 run average), and
the size of the data structure representing the dataset. Values were
obtained by running scripts in \texttt{benchmarks} on Apple M2
Max CPU.}
\vspace{-18pt}
\begin{longtable}[]{@{}
  >{\raggedright\arraybackslash}p{(\columnwidth - 8\tabcolsep) * \real{0.1478}}
  >{\raggedright\arraybackslash}p{(\columnwidth - 8\tabcolsep) * \real{0.1478}}
  >{\centering\arraybackslash}p{(\columnwidth - 8\tabcolsep) * \real{0.1652}}
  >{\centering\arraybackslash}p{(\columnwidth - 8\tabcolsep) * \real{0.2609}}
  >{\centering\arraybackslash}p{(\columnwidth - 8\tabcolsep) * \real{0.2522}}@{}}
\tabularnewline
\toprule\noalign{}
\begin{minipage}[b]{\linewidth}\raggedright
Tool
\end{minipage} & \begin{minipage}[b]{\linewidth}\raggedright
Object
\end{minipage} & \begin{minipage}[b]{\linewidth}\centering
Time per Dataset
\end{minipage} & \begin{minipage}[b]{\linewidth}\centering
Time per Entry \emph{(Relative)}
\end{minipage} & \begin{minipage}[b]{\linewidth}\centering
Database Size \emph{(Relative)}
\end{minipage} \\
\midrule\noalign{}
\endfirsthead
\toprule\noalign{}
\begin{minipage}[b]{\linewidth}\raggedright
Tool
\end{minipage} & \begin{minipage}[b]{\linewidth}\raggedright
Object
\end{minipage} & \begin{minipage}[b]{\linewidth}\centering
Time per Dataset
\end{minipage} & \begin{minipage}[b]{\linewidth}\centering
Time per Entry \emph{(Relative)}
\end{minipage} & \begin{minipage}[b]{\linewidth}\centering
Database Size \emph{(Relative)}
\end{minipage} \\
\midrule\noalign{}
\endhead
\midrule\noalign{}
\texttt{Python}\textsuperscript{3.11} & \texttt{set} & 327.4 µs & 152.3
ns \emph{(x1)} & 871.5 kB \emph{(x1)} \\
\texttt{NumPy}\textsuperscript{1.26} & \texttt{array} & 40.1 µs & 18.6
ns \emph{(x8.3)} & 79.7 kB \emph{(x10.9)} \\
\texttt{nimCSO}\textsuperscript{0.6} & \texttt{BitArray} & 9.2 µs & 4.4
ns \emph{(x34.6)} & 50.4 kB \emph{(x17.3)} \\
\texttt{nimCSO}\textsuperscript{0.6} & \texttt{uint64} & 0.79 µs & 0.37
ns \emph{(x413)} & 16.8 kB \emph{(x52)} \\
\bottomrule\noalign{}
\endlastfoot
\end{longtable}

\hypertarget{brute-force-search}{%
\subsection{Brute-Force Search}\label{brute-force-search}}

The brute-force search is a naïve method of evaluating all
possibilities; however, its near-zero overhead can make it the most
efficient for small problems. In this implementation, all entries in the
\emph{power set} of \(N\) considered elements are represented as a range
of integers from \(0\) to \(2^{N} - 1\), and used to initialize
\texttt{uint64}/\texttt{BitArray}s on the fly. To minimize the memory
footprint of solutions, the algorithm only keeps track of the best
solution for a given number of elements present in the solution. Current
implementations are limited to 64 elements, as it is not feasible beyond
approximately 30 elements; however, the one based on \texttt{BitArray}
could be easily extended if needed.

\hypertarget{algorithm-based-search}{%
\subsection{Algorithm-Based Search}\label{algorithm-based-search}}

The algorithm implemented in the \texttt{algorithmSearch} routine,
targeting high dimensional problems (20-50), iteratively expands and
evaluates candidates from a priority queue (implemented through an
efficient binary heap (\protect\hyperlink{ref-Williams1964}{Williams,
1964})) while leveraging the fact that \emph{the number of data points
lost when removing elements \texttt{A} and \texttt{B} from the dataset
has to be at least as large as when removing either \texttt{A} or
\texttt{B} alone} to delay exploration of candidates until they can
contribute to the solution. Furthermore, to (1) avoid revisiting the
same candidate without keeping track of visited states and (2) further
inhibit the exploration of unlikely candidates, the algorithm
\emph{assumes} that while searching for a given order of solution,
elements present in already expanded solutions will not improve those
not yet expanded. This effectively prunes candidate branches requiring
two or more levels of backtracking. In the authors' tests, this method
has generated the same results as \texttt{bruteForce}, except for
occasional differences in the last explored solution.

\hypertarget{genetic-search}{%
\subsection{Genetic Search}\label{genetic-search}}

\begin{figure}[h]
  \centering
  \includegraphics[width=1.31944in,height=\textheight]{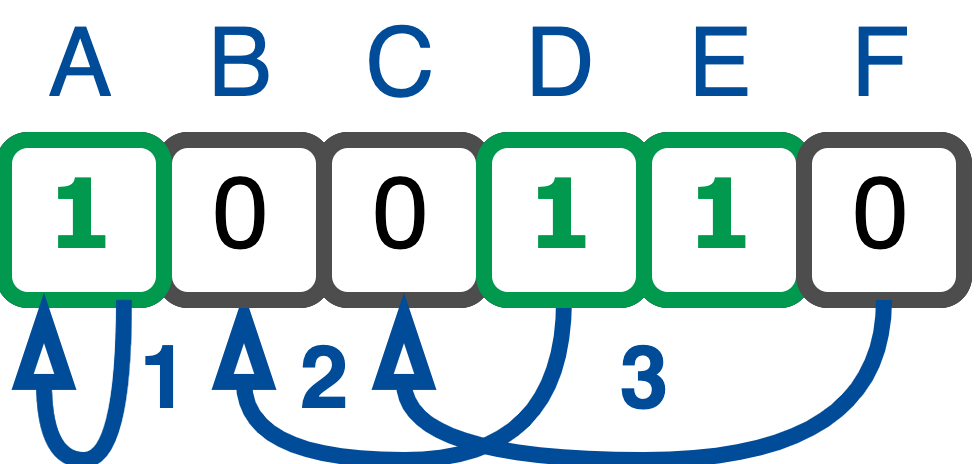}
  \caption{Schematic of \texttt{mutate} procedure where bits are swapping
  randomly, so that (1) bit can swap itself, (2) bits can swap causing a
  flip, or (3) bits can swap with no effect.}
  \label{fig:mutate}
\end{figure}

Beyond 50 components, the
\protect\hyperlink{algorithm-based-search}{algorithm-based} method will
likely run out of memory on most personal systems. The
\texttt{geneticSearch} routine resolves this issue through an evolution
strategy to iteratively improve solutions based on custom
\texttt{mutate} and \texttt{crossover} procedures. Both are of uniform
type (\protect\hyperlink{ref-Goldberg1989}{Goldberg, 1989}) with
additional constraint of Hamming weight
(\protect\hyperlink{ref-Knuth}{Knuth, 2009}) preservation in order to
preserve number of considered elements in parents and offspring. In
\texttt{mutate} this is achieved by using purely random bit swapping,
rather than more common flipping, as demonstrated in the Figure
\ref{fig:mutate}.

\begin{figure}[h]
\centering
\includegraphics[width=4.16667in,height=\textheight]{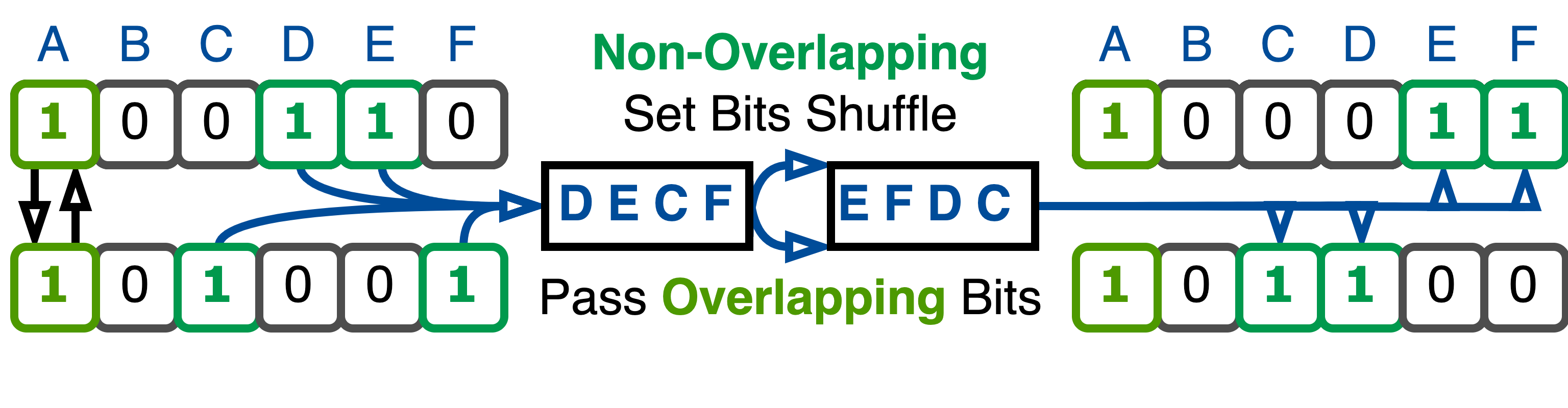}
\caption{Schematic of uniform \texttt{crossover} procedure preserving
Hamming weight implemented in \texttt{nimCSO}.}
\label{fig:crossover}
\end{figure}

Meanwhile, in \texttt{crossover}, this constraint is satisfied by
passing overlapping bits directly, while non-overlapping bits are
shuffled and distributed at positions present in one of the parents, as
shown in Figure \ref{fig:crossover}.

The above are applied iteratively, with best solutions carried to next
generation, until the solution converges or the maximum number of
iterations is reached. Unlike the other methods, the present method is
not limited by the number of components and lets user control both time
and memory requirements, either to make big problems feasible or to get
a good-enough solution quickly in small problems. However, it comes with
no optimality guarantees.

\hypertarget{use-examples}{%
\section{Use Examples}\label{use-examples}}

The tool comes with two pre-defined example problems to demonstrate its
use. The first one is defined in the default \texttt{config.yaml} file
and goes through the complete dataset of 2,150 data points spanning 37
components in \texttt{dataList.txt} based on the ULTERA Dataset
(\protect\hyperlink{ref-Debnath2021}{Debnath et al., 2021}). It is
intended to showcase \texttt{algorithmSearch}/\texttt{-as} and
\texttt{geneticSearch}/\texttt{-gs} methods, as brute-forcing would take
around one day. The second one is defined in \texttt{config\_rhea.yaml}
and uses the same dataset but a limited scope of components critical to
RHEAs (\protect\hyperlink{ref-Senkov2018}{Senkov et al., 2018}) and is
intended to showcase \texttt{bruteForce}/\texttt{-bf} and
\texttt{bruteForceInt}/\texttt{-bfi} methods. With four simple commands
(see Table 2), the user can compare the methods' performance and the
solutions' quality.

\begin{longtable}[]{@{}
  >{\raggedright\arraybackslash}p{(\columnwidth - 4\tabcolsep) * \real{0.6623}}
  >{\raggedright\arraybackslash}p{(\columnwidth - 4\tabcolsep) * \real{0.1429}}
  >{\centering\arraybackslash}p{(\columnwidth - 4\tabcolsep) * \real{0.1818}}@{}}
\caption{Four example tasks alongside typical CPU time and memory usage
on Apple M2 Max.}\tabularnewline
\toprule\noalign{}
\begin{minipage}[b]{\linewidth}\raggedright
Task Definition (\texttt{nim\ c\ -r\ -f\ -d:release\ ...})
\end{minipage} & \begin{minipage}[b]{\linewidth}\raggedright
Time (s)
\end{minipage} & \begin{minipage}[b]{\linewidth}\centering
Memory (MB)
\end{minipage} \\
\midrule\noalign{}
\endfirsthead
\toprule\noalign{}
\begin{minipage}[b]{\linewidth}\raggedright
Task Definition (\texttt{nim\ c\ -r\ -f\ -d:release\ ...})
\end{minipage} & \begin{minipage}[b]{\linewidth}\raggedright
Time (s)
\end{minipage} & \begin{minipage}[b]{\linewidth}\centering
Memory (MB)
\end{minipage} \\
\midrule\noalign{}
\endhead
\bottomrule\noalign{}
\endlastfoot
\texttt{-d:configPath=config.yaml\ src/nimcso\ -as} & 302s & 488 MB \\
\texttt{-d:configPath=config.yaml\ src/nimcso\ -gs} & 5.8s & 3.2 MB \\
\texttt{-d:configPath=config\_rhea.yaml\ src/nimcso\ -as} & 0.076s & 2.2
MB \\
\texttt{-d:configPath=config\_rhea.yaml\ src/nimcso\ -gs} & 0.429s & 2.1
MB \\
\texttt{-d:configPath=config\_rhea.yaml\ src/nimcso\ -bf} & 4.171s & 2.0
MB \\
\texttt{-d:configPath=config\_rhea.yaml\ src/nimcso\ -bfi} & 0.459s &
2.0 MB \\
\end{longtable}

In case of issues, the help message can be accessed by running the tool
with \texttt{-h} flag or by refering to documentation at
\href{https://nimcso.phaseslab.org}{nimcso.phaseslab.org}.

\hypertarget{contributions}{%
\section{Contributions}\label{contributions}}

A.M.K. was reponsible for conceptualization, methodology, software,
testing and validation, writing of manuscript, and visualization; A.D.
was responsible for testing software and results in training machine
learning models; A.M.B., W.F.R., Z-K.L. were responsible for funding
acquisition, review, and editing. Z-K.L. was also supervising the work.

\hypertarget{acknowledgements}{%
\section{Acknowledgements}\label{acknowledgements}}

This work has been funded through grants: \textbf{DOE-ARPA-E
DE-AR0001435}, \textbf{NSF-POSE FAIN-2229690}, and \textbf{ONR
N00014-23-2721}. We would also like to acknowledge Dr.~Jonathan Siegel
at Texas A\&M University for several valuable discussions and feedback
on the project.

\hypertarget{references}{%
\section*{References}\label{references}}
\addcontentsline{toc}{section}{References}

\hypertarget{refs}{}
\begin{CSLReferences}{1}{0}
\leavevmode\vadjust pre{\hypertarget{ref-Cantor2004}{}}%
Cantor, B., Chang, I. T. H., Knight, P., \& Vincent, A. J. B. (2004).
Microstructural development in equiatomic multicomponent alloys.
\emph{Materials Science and Engineering A}, \emph{375-377}, 213--218.
\url{https://doi.org/10.1016/j.msea.2003.10.257}

\leavevmode\vadjust pre{\hypertarget{ref-Debnath2021}{}}%
Debnath, A., Krajewski, A. M., Sun, H., Lin, S., Ahn, M., Li, W., Priya,
S., Singh, J., Shang, S., Beese, A. M., Liu, Z.-K., \& Reinhart, W. F.
(2021). \emph{Journal of Materials Informatics}, \emph{1}.
\url{https://doi.org/10.20517/jmi.2021.05}

\leavevmode\vadjust pre{\hypertarget{ref-Evans2024}{}}%
Evans, M. L., Bergsma, J., Merkys, A., Andersen, C. W., Andersson, O.
B., Beltrán, D., Blokhin, E., Boland, T. M., Balderas, R. C., Choudhary,
K., Díaz, A. D., García, R. D., Eckert, H., Eimre, K., Montero, M. E.
F., Krajewski, A. M., Mortensen, J. J., Duarte, J. M. N., Pietryga, J.,
\ldots{} Armiento, R. (2024). \emph{Developments and applications of the
OPTIMADE API for materials discovery, design, and data exchange}.
\url{https://doi.org/10.48550/arXiv.2402.00572}

\leavevmode\vadjust pre{\hypertarget{ref-Goldberg1989}{}}%
Goldberg, D. E. (1989). \emph{Genetic algorithms in search, optimization
and machine learning} (1st ed.). Addison-Wesley Longman Publishing Co.,
Inc. ISBN:~0201157675

\leavevmode\vadjust pre{\hypertarget{ref-Knuth}{}}%
Knuth, D. E. (2009). \emph{The art of computer programming, volume 4,
fascicle 1: Bitwise tricks \& techniques; binary decision diagrams}
(12th ed.). Addison-Wesley Professional. ISBN:~0321580508

\leavevmode\vadjust pre{\hypertarget{ref-Krajewski2024Nimplex}{}}%
Krajewski, A. M., Beese, A. M., Reinhart, W. F., \& Liu, Z.-K. (2024).
\emph{Efficient generation of grids and traversal graphs in
compositional spaces towards exploration and path planning exemplified
in materials}. \url{http://arxiv.org/abs/2402.03528}

\leavevmode\vadjust pre{\hypertarget{ref-Li2024}{}}%
Li, W., Raman, L., Debnath, A., Ahn, M., Lin, S., Krajewski, A. M.,
Shang, S., Priya, S., Reinhart, W. F., Liu, Z.-K., \& Beese, A. M.
(2024). \emph{Design and validation of refractory alloys using machine
learning, CALPHAD, and experiments}.
https://doi.org/\url{https://dx.doi.org/10.2139/ssrn.4689687}

\leavevmode\vadjust pre{\hypertarget{ref-Rumpf2023}{}}%
Rumpf, A. (2023). \emph{Nim programming language v2.0.0}.
\url{https://nim-lang.org/}

\leavevmode\vadjust pre{\hypertarget{ref-Senkov2018}{}}%
Senkov, O. N., Miracle, D. B., Chaput, K. J., \& Couzinie, J.-P. (2018).
Development and exploration of refractory high entropy alloys---a
review. \emph{Journal of Materials Research}, \emph{33}, 3092--3128.
\url{https://doi.org/10.1557/jmr.2018.153}

\leavevmode\vadjust pre{\hypertarget{ref-Williams1964}{}}%
Williams, J. W. J. (1964). Algorithm 232 - heapsort.
\emph{Communications of the ACM}, \emph{7}, 347--349.
\url{https://doi.org/10.1145/512274.512284}

\leavevmode\vadjust pre{\hypertarget{ref-Yeh2004}{}}%
Yeh, J. W., Chen, S. K., Lin, S. J., Gan, J. Y., Chin, T. S., Shun, T.
T., Tsau, C. H., \& Chang, S. Y. (2004). Nanostructured high-entropy
alloys with multiple principal elements: Novel alloy design concepts and
outcomes. \emph{Advanced Engineering Materials}, \emph{6}, 299--303.
\url{https://doi.org/10.1002/adem.200300567}

\end{CSLReferences}

\end{document}